\documentclass{article}
\usepackage[utf8]{inputenc}

\usepackage{cite}
\usepackage{amsmath,amssymb,amsfonts}
\usepackage{algorithm}
\usepackage{algorithmic}
\usepackage{graphicx}
\usepackage{array}
\usepackage{textcomp}
\usepackage{xcolor}
\usepackage{fullpage}
\usepackage{caption}
\usepackage{subcaption}
\usepackage{float}
\usepackage{comment}
\usepackage{multirow}
\usepackage{footnote}
\usepackage{csquotes}
\usepackage{hyperref}
\usepackage{authblk}

\newcolumntype{P}[1]{>{\centering\arraybackslash}p{#1}}

\def\BibTeX{{\rm B\kern-.05em{\sc i\kern-.025em b}\kern-.08em
    T\kern-.1667em\lower.7ex\hbox{E}\kern-.125emX}}

\providecommand{\keywords}[1]
{
  \small	
  \textbf{\textit{Keywords---}} #1
}

\title{AI-based Malware and Ransomware\\Detection Models}

\author[1,2]{Benjamin Marais}
\author[1]{Tony Quertier}
\author[1]{Stéphane Morucci}

\affil[1]{Orange Innovation, Rennes, France\thanks{\texttt{[benjamin.marais, tony.quertier,
stephane.morucci]@orange.com}}}
\affil[2]{Department of Mathematics, LMNO, University of Caen Normandy, France}

\date{}

\begin{document}

\maketitle

\begin{abstract} Cybercrime is one of the major digital threats of this century. In particular, ransomware attacks have significantly increased, resulting in global damage costs of tens of billion dollars. In this paper, we train and test different Machine Learning and Deep Learning models for malware detection, malware classification and ransomware detection. We introduce a novel and flexible solution that combines two optimized models for malware and ransomware detection. Our results demonstrate some improvements both in terms of detection performances and flexibility. In particular, our combined models pave the way for easier future enhancements using specialized and thus interchangeable detection modules.
\end{abstract}


\section{Introduction}

The cybercrime economy has never been so lucrative. In 2021, the global cost of cybercrime campaigns damages reached about 6 trillion USD for individuals and companies \cite{cybercrimemag}. The trend is not expected to change since the estimated cost by 2025 is about 10.5 trillion USD. Since the beginning of the COVID-19 pandemic, cyber-attack campaigns have multiplied \cite{Lallie2021} and according to AV-Test \cite{avtest}, 150 million new malicious files have been discovered in 2021, an increase of $36\%$ compared to 2020. Malware detection is therefore a major issue for individuals, companies and even governments. Recently, Republic of Costa Rica had to declare a state of national emergency due to a ransomware attack carried out by Conti hacker group \cite{emergency-Costa-Rica}. Despite the progress made in malware and ransomware detection, the problem remains and intensifies over time, mainly because hackers techniques are constantly and rapidly evolving. To perform faster remediation activities, it is very important for security teams to quickly identify the family or the category of a detected malware. This classification problem can be successfully addressed using Machine Learning (ML) and Deep Learning (DL) techniques as soon as enough labeled data are available. In this work, we are training different algorithms to detect malware and ransomware. In particular, we build a bi-layered ransomware detection model based on two ML and DL optimized models. Static techniques are used to extract prevalent features from Windows Portable Executable (PE) files to train our models.

\subsection{Background and Related Work}

Malicious software analysis is a major research topic due to the damages malware cause \cite{Anderson2019MeasuringTC}. With the recent advances in Artificial Intelligence (AI), cybersecurity researchers are shifting their attention to Machine Learning  (ML) and Deep Learning (DL) methods to improve malicious files detection and classification \cite{Ucci2019, Raff2020, Gibert2020}. Even if results are decent, these methods still need improvements \cite{Smith2020}. 

Another hot research area in malware analysis relates to detection of a particular malicious file type, like a ransomware for instance. Ransomware aim at disabling the functionality of a computer, either by encrypting the machine (cryptographic-ransomware) as done by the well-known Wannacry virus \cite{Mohurle2017, Y.Connolly2019}, or by blocking access to the machine (locker-ransomware) as performed by Reveton malware \cite{Kharraz2015}. To regain the control of a computer, malware authors usually require to pay a ransom. Although this threat has been around for decades, it has intensified with the rise of cryptocurrencies which make it possible to receive a payment with a certain level of anonymity. Even though defensive methods exist to prevent such attack \cite{Richardson2017, Kolodenker2017, Oz2022, Maniath2019, McAfee}, $54\%$ of them succeeded in $2021$ at an estimated average unitary cost of $1.85$ million USD.  \cite{SophosLtd2021}.

\subsection{Contributions}

This study aims to provide different approaches for malware detection with a particular focus on ransomware detection. Our main contribution is an algorithm that chains two models optimized for malicious files detection and ransomware detection respectively. This solution, called "bi-layered detection model", takes as input a PE file, outputs the maliciousness of the file in a second step, and then determines if this PE file can be classified as a ransomware or not. To the best of our knowledge, we didn't find any similar approach for ransomware detection in the existing literature. Our final model is thus composed of two specialized models that can be independently trained which makes it very flexible during optimization phases and in a production environment.

\subsection{Outline}

In Section \ref{sec:DataFeatures}, we present datasets and feature extraction methods used for our experiments. In Section \ref{sec:Detection}, we train the model for the first step: malicious file detection. In Section \ref{sec:class}, we create a model for the classification of malware families and in particular of the ransomware category. In Section \ref{sec:allin}, we train a bi-layered model that performs malware and ransomware detection and compare it to a reference model. Finally, Section \ref{sec:conclusion} summarizes this paper and discusses future works. 

\section{Datasets, Features and Models}
\label{sec:DataFeatures}
\subsection{Datasets}
\label{Sec:Dataset}

For our experiments, we rely on three different datasets. They all contain malicious and benign Portable Executable (PE) files in two different formats. These datasets are Ember \cite{Anderson2018}, Bodmas \cite{Yang} and PEMachineLearning \cite{web1}. PE files distribution and format are summarized in Table \ref{tab:dataset}.

\begin{table}[!ht]
    \centering
    \caption{Distribution and format of each dataset}
    \label{tab:dataset}
    \begin{tabular}{|P{.09\textwidth}|P{.09\textwidth}|P{.09\textwidth}|P{.09\textwidth}|}
        \hline
         & Malicious files & Benign files & Files format  \\
         \hline
        Ember & 400,000 & 400,000 & Features\\
        \hline
        \multirow{2}{*}{Bodmas} &57,293 & 77,142 & Features\\
        \cline{2-4}
        & 57,293 & 0 & PE files\\
        \hline
        PEMachine Learning & 114,737 & 86,812 & PE files\\
        \hline
    \end{tabular}
\end{table}

\subsubsection*{Ember}

Ember is a dataset provided by Anderson et al. \cite{Anderson2018}. It contains a total of 1.1M features extracted from 400K malicious files, 400K benign files and 300K unlabeled files. Anderson et al. also provide the necessary tooling to generate a feature-based dataset. 

\subsubsection*{Bodmas}

Bodmas \cite{Yang} shared with our team a dataset that contains $134,435$ binary files in the same format as Ember with pre-extracted features together with the $57,293$ malicious files in raw PE format. These files have been collected during one year between August 2019 and September 2020 and labeled: authors indicate the category each file belongs to. Table \ref{tab:Bodmas} presents the distribution of malware families in the Bodmas dataset. The category "other" includes about ten other categories less represented in the Bodmas dataset such as "dropper", "downloader" or "informationstealer" for instance.

\begin{table}[!ht]
    \begin{center}
    \caption{Distribution of malware families in the Bodmas dataset}
    \label{tab:Bodmas}
    \begin{tabular}{|c|c|}
        \hline
         Category & Files count\\
         \hline
         Trojan & 29,972 \\
         Worm & 16,697 \\
         Backdoor & 7,331 \\
         Ransomware & 821 \\
         Other & 2,471 \\
         \hline
         Total & 57,293 \\
         \hline
    \end{tabular}
    \end{center}
\end{table}

\subsubsection*{PEMachineLearning Dataset}

The third dataset used is PEMachineLearning, made available by M. Lester \cite{web1}. It contains $201,549$ binary files including $114,737$ malicious files. These files have been gathered from different sources such as VirusShare\footnote{\url{https://virusshare.com/}}, MalShare\footnote{\url{https://malshare.com/}} and TheZoo\footnote{\url{https://github.com/ytisf/theZoo}}.  

\subsection{Features}

\begin{figure}[!ht]
    \centering
    \begin{subfigure}[!ht]{0.2\textwidth}
        \includegraphics[width=\textwidth]{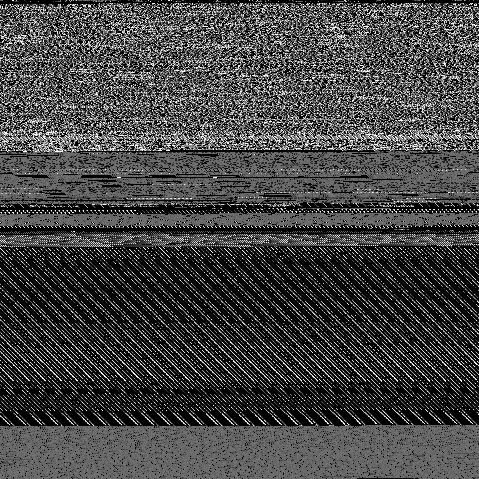}
        \caption{Benign file}
        \label{fig:grayscale_1}
    \end{subfigure}
    \begin{subfigure}[!ht]{0.2\textwidth}
        \includegraphics[width=\textwidth]{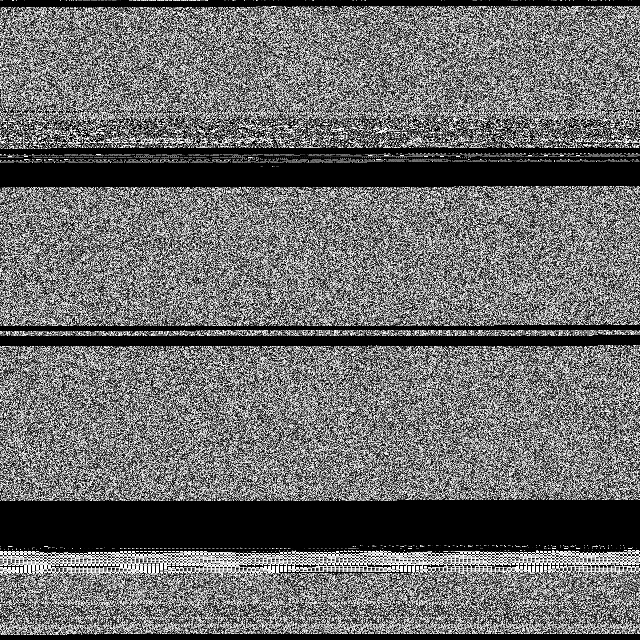}
        \caption{Malicious file}
        \label{fig:grayscale_2}
    \end{subfigure}
        \begin{subfigure}[!ht]{0.2\textwidth}
        \includegraphics[width=\textwidth]{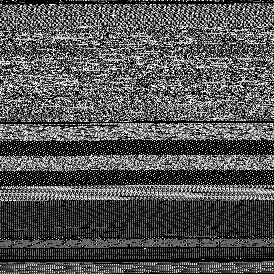}
        \caption{Ransomware file}
        \label{fig:grayscale_3}
    \end{subfigure}
    \caption{Example of PE files transformation into grayscale images}
    \label{fig:grayscale}
\end{figure}

Training models for malware detection is a multi-step process. The first step consists in extracting information (i.e. features) from the PE files. For this, we rely on two pre-processing algorithms. The first one, a.k.a the Ember method, is detailed in Anderson et al. \cite{Anderson2018} and converts a PE file into a vector of $2,381$ features. Some of these features are listed in Table \ref{tab:Ember}. The second one, a.k.a Grayscale method, was initially submitted by Nataraj et al. \cite{Nataraj2011} and converts a binary file into an image, as can be seen in Figure \ref{fig:grayscale}. To train our models, we choose to resize grayscale images $64\times64$ pixels.

\begin{table}[!ht]
    \begin{center}
    \caption{Some features computed using Ember extractor}
    \label{tab:Ember}
    \begin{tabular}{|c|c|}
        \hline
         Feature names & Index\\
         \hline
         Byte histogram & 1-256 \\
         Byte entropy & 257-512 \\
         Strings  & 513-616 \\
         General information & 617-626 \\
         Header information & 627-688 \\
         Section & 689-943 \\
         Imports & 944-2223 \\
         Exports & 2224-2350 \\
         Data directory information & 2252 - 2381 \\
         \hline
    \end{tabular}
    \end{center}
\end{table}

\subsection{Models}

We selected four machine learning (ML) and deep learning (DL) models:  

\begin{itemize}
    \item Three models are trained with features extracted from PE files using the Ember extractor.  
    \begin{itemize}
        \item LightGBM, 
        \item XGBoost,
        \item Dense Neural Network (DNN),
    \end{itemize}
    \item a Convolution Neural Network (CNN) trained with PE files images created by the Grayscale extractor.
\end{itemize}

For LightGBM and XGBoost models, we use the default parameters provided by the two python packages. For DNN and CNN models, we use Tensorflow package with the Adam optimization function \cite{Kingma2015} and a learning rate of $0.01$. 

\section{Detection of malicious files}
\label{sec:Detection}

In this section, we are discussing the detection of malicious software using different models of machine learning. For this, we use PEMachineLearning and then Bodmas dataset. PEMachineLearning has been divided into three subsets for training ($70\%$), validation ($15\%$) and testing ($15\%$) of our models. 

The four models are first trained and tested on PEMachineLearning. We compare the models results using the F1 score and the accuracy score. Table \ref{tab:detection_1} presents the results on the test subset. The algorithm XGboost has the best results although the performance of LightGBM and DNN are relatively close. We observe that the CNN is less efficient than the three other models even if scores are close to $0.95$.

\begin{table}[!ht]
    \begin{center}
        \caption{Malware detection results on the PEMachineLearning test subset}
        \label{tab:detection_1}
        \begin{tabular}{|c|c|c|c|c|}
            \hline
             & LGBM & XGBoost & DNN & CNN \\ 
             \hline
             Accuracy & 0.990 & \textbf{0.993} &  0.9902 & 0.9458 \\ 
             \hline
             F1 Score & 0.991 & \textbf{0.994} & 0.991 & 0.95 \\
             \hline
        \end{tabular}
    \end{center}
\end{table}

Subsequently, we test our model on malicious files from Bodmas dataset: the purpose is to validate the robustness of our trained models and to determine if they do not produce too many false negatives predictions i.e. true malware detected as benign. We focus on the false negative rate (FNR) as we consider it as a prevalent metric. For each model, FNR is showed in Table \ref{tab:detection_2}. We also add the number of undetected malware out of 57,293 files from Bodmas. The CNN model has a high FNR compared to the other models: it does not detect enough malware from Bodmas, meaning that this model does not have a good generalization capacity. On the contrary, the XGBoost and LightGBM models have very close and low FNRs. They achieve good performances during the testing phase and have good generalization capacities. Finally, the DNN has a perfect score of zero undetected malware. 

We compare our results on Bodmas with a current state of the art model. We test the LightGBM model provided by Ember \cite{Anderson2018} on the Bodmas dataset and it achieves a FNR of $1.42\cdot10^{-2}$. With the exception of the CNN, the models we propose are slightly better than Ember's model, even though they are trained with less but more recent data. 

\begin{table}[!ht]
    \begin{center}
        \caption{FNR and number of undetected malware from Bodmas dataset}
        \label{tab:detection_2}
        \begin{tabular}{|c|c|c|c|c||c|}
            \hline
             & LGBM & XGBoost & DNN & CNN & Ember LGBM\\ 
             \hline
            FNR & $1.56\cdot10^{-3}$& $0.96\cdot10^{-3}$ & \textbf{0} & 0.13 & $1.42\cdot10^{-2}$\\
             \hline
            Undetected malware & 84 & 55 & \textbf{0} & 7448 & 816\\
            \hline
        \end{tabular}
    \end{center}
\end{table}

Given results from Table \ref{tab:detection_1} and Table \ref{tab:detection_2}, the XGBoost model seems to be the most effective model for detecting malicious files during training and testing phases. Moreover, it shows a good capacity of generalization with a low FNR on Bodmas dataset. LightGBM model provides slightly lower but very close results. Its main advantage over XGBoost is its faster computing time \cite{Ke2017}. Even though results for the CNN model are quite good, its poor FNR performance indicates a low generalization capacity. Finally, the DNN seems to perform better than others with results close to XGBoost on the testing subset and the lowest FNR, equal to zero, on the Bodmas dataset. On the other hand, it requires more computing power for a rather small performance increase. Thus, XGBoost model could be considered as a good trade-off between computing time and performance.

\section{Malware Categories Classification and Ransomware Detection}
\label{sec:class}

The purpose of this section is to present some experiments and results in the context of malware classification. We pursue two objectives: the first one consists in identifying four of the most popular malware categories (cf Section \ref{sec:clas_5lab}). The second objective aims at detecting ransomware only and results are presented in Section \ref{sec:class_rsw}. In both cases, we use the same two extractors and the same models as presented in Section \ref{sec:DataFeatures}. 

Regarding the datasets at our disposal, we mainly use Bodmas because it provides labeled malicious files. However, due to a lack of ransomware, we had to manually labeled several malware from PEMachineLearning to reach a total of 2,000 labeled files. 

\subsection{Malware classification}
\label{sec:clas_5lab}

In this part, we are classifying malware into four popular and frequently encountered malware's categories i.e. Trojan, Worm, Backdoor and Ransomware. All other types of malicious files fall into the "Other" category. To limit overfitting, we decided to use a balanced dataset instead of using all available data at our disposal. Thus, for each category of malware, we selected 2,000 malicious files from Bodmas and PEMachileLearning dataset (for ransomware). This balanced dataset has been split into a training subset (70\%), a validation subset (15\%) and a test subset (15\%). 

To compare and determine which model is the most effective for malware classification, we rely on the accuracy and F1 scores, summarized in Table\ref{tab:5_classification}. The best results are obtained using LightGBM, even if XGBoost and DNN performances are really close. With an average accuracy score of $0.9413$ and an average F1 score of $0.9412$, these three models appear to be good candidates for malware classification even if these classification results could be improved.  

\begin{table}[!ht]
    \begin{center}
        \caption{Malware classification results ont the test subset}
        \label{tab:5_classification}
        \begin{tabular}{|c|c|c|c|c|}
            \hline
             & LGBM & XGBoost & DNN & CNN \\ 
             \hline
             Accuracy & \textbf{0.9442} & 0.9427 &  0.9369 & 0.7270 \\ 
             \hline
             F1 Score & \textbf{0.9440} & 0.9425 & 0.9370 & 0.7197 \\
             \hline
        \end{tabular}
    \end{center}
\end{table}

On the contrary, CNN doesn't perform as good as other models and its results cannot be considered as reliable. According to us, this is not only due to a low number of samples during training phase, but also to a sub-optimized pre-processing step.

\subsection{Ransomware Detection}
\label{sec:class_rsw}

In this section, we are focusing on detecting ransomware among other malicious files. For this purpose, we train models on the same dataset as in Section \ref{sec:clas_5lab}. It is composed of 2,000 ransomware files and 8,000 malicious files. To compare the four models trained for ransomware detection, we use F1 and accuracy scores as metrics, summarized in Table \ref{tab:2_class}. LightGBM, XGBoost and DNN achieve almost the same performance with scores close to $1$ on the test subset. For this classification task, CNN also achieves good results.

\begin{table}[!ht]
    \begin{center}
        \caption{Ransomware detection results on the test subset}
        \label{tab:2_class}
        \begin{tabular}{|c|c|c|c|c|}
            \hline
             & LGBM & XGBoost & DNN & CNN \\ 
             \hline
             Accuracy & \textbf{0.9954} & 0.9948 &  0.9938 & 0.9830 \\ 
             \hline
             F1 Score & \textbf{0.9971} & 0.9969 & 0.9967 & 0.9823 \\
             \hline
        \end{tabular}
    \end{center}
\end{table}

Once again, LGBM performs better even if DNN and XGBoost are also good candidates. Our results demonstrate that it seems easier to separate the ransomware category from others than to classify malware into five categories as done in Section \ref{sec:clas_5lab}. 

\section{Bi-Layered Model for Malware and Ransomware Detection}
\label{sec:allin}

\begin{figure}[!ht]
    \centering
    \begin{subfigure}[h]{0.45\textwidth}
        \includegraphics[width=\textwidth]{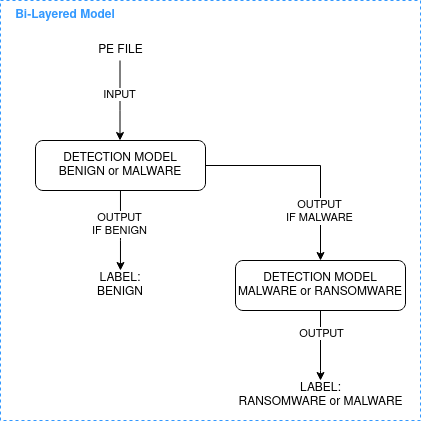}
        \caption{Bi-Layered Model}
        \label{fig:models_1}
    \end{subfigure}
    \begin{subfigure}[h]{0.45\textwidth}
        \includegraphics[width=\textwidth]{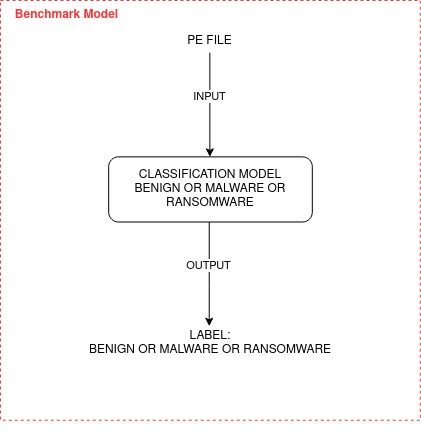}
        \caption{Benchmark Model}
        \label{fig:model2}
    \end{subfigure}
    \caption{Ransomware models diagrams}
    \label{fig:models}
\end{figure}

In this section, we are considering an algorithm referred to as \textquote{bi-layered detection model}. Its purpose is to detect PE files that are malicious, and also if they belong to the ransomware category. This algorithm consists in two detection layers. The first one relies on a model from Section \ref{sec:Detection} and its purpose is to detect malicious files. The second layer uses a model trained for ransomware detection given a malicious file (cf Section \ref{sec:class_rsw}). We are combining them as presented in Figure \ref{fig:models_1} to predict if a file is benign and if it can be classified as a ransomware or not.
To evaluate the performances of this combined model, we are training several reference models, referred to as \textquote{benchmark models}. They are based on LightGBM, XGBoost and DNN using dataset from Section \ref{sec:class_rsw} and $8,000$ benign files from PEMachineLearning. 

\begin{figure}[ht]
    \centering
    \begin{subfigure}[h]{0.45\textwidth}
        \includegraphics[width=\textwidth]{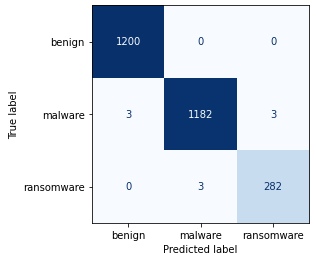}
        \caption{Bi-layered XGBoost}
        \label{fig:matrix1}
    \end{subfigure}
    \begin{subfigure}[h]{0.45\textwidth}
        \includegraphics[width=\textwidth]{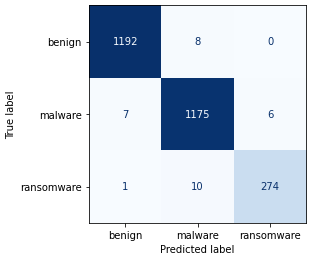}
        \caption{Benchmark XGBoost}
        \label{fig:matrix2}
    \end{subfigure}
    \caption{Confusion matrix of bi-layered and benchmark XGBoost on the test subset}
    \label{fig:matrix}
\end{figure}

For each model, Table \ref{tab:blocks} summarizes the accuracy and F1 scores. Firstly, we see that our bi-layered detection models have slightly better scores than benchmark models. Overall, bi-layered detection models appear to perform better than benchmark models. We also present confusion matrices of the bi-layered and benchmark XGboost models in Figure \ref{fig:matrix}. We can notice that bi-layered XGBoost returns less misclassified PE files than benchmark, which confirms that our bi-layered detection models seem to be more accurate. Finally, the XGBoost bi-layered model has a better efficiency than DNN and LightGBM, even if results are close.

Moreover, a significant advantage of our bi-layer detection models is that they are composed of two sub-models trained for particular sub-tasks i.e. malware or ransomware detection. In consequence, each sub-model can be retrained independently if we have new data, and possibly be replaced if a better model becomes available. We are convinced that such flexibility makes our models good candidates for production environment. 

\begin{table}[h]
    \caption{Results for Benchmark and Bi-Layered models on the test subset}
    \label{tab:blocks}
    \centering
    \begin{tabular}{{|P{.1\textwidth}|P{.1\textwidth}|P{.1\textwidth}|P{.1\textwidth}|P{.1\textwidth}|}}
    \hline
    \multicolumn{1}{|c|}{} & 
    \multicolumn{2}{c|}{Accuracy} &
    \multicolumn{2}{c|}{F1 Score } \\
    \hline
     & Benchmark & Bi-Layered & Benchmark & Bi-Layered \\
    \hline
    LGBM & 0.9896 & 0.9932 & 0.9895 & 0.9934 \\ 
    \hline
    XGBoost & 0.988 & \textbf{0.9960} & 0.988 &  \textbf{0.9966} \\
    \hline
    DNN & 0.9867 & 0.9913 & 0.9868 & 0.9915 \\
    \hline
    \end{tabular}
\end{table}

\section{Conclusion \& Future Work}
\label{sec:conclusion}

\subsection{Conclusion}

In this paper, we have implemented and compared different methods, based on machine learning and deep learning algorithms, for malware detection and classification, and in particular ransomware detection. During each experimentation, CNN models do not perform as well as other models. We need to investigate further to identify the origin of the problem, even if we feel that the lack of data or prepossessing steps could be an explanation. 
In Section \ref{sec:Detection}, all other models achieve good results, in particular XGBoost which is the most effective on the test subset and DNN which seems to have better generalization capacity. Thus, XGBoost may appear as a good candidate for malware detection considering their respective training times. 
In Section \ref{sec:clas_5lab}, we show that about $5\%$ of PE files are misclassified which leaves room for models optimizations. In Section \ref{sec:class_rsw}, by focusing on ransomware detection only, we demonstrate that XGBoost, LightGBM and DNN have good performances with an average accuracy score of $0.9947$. It coud be interesting to analyze if these findings are similar for other category of malicious files like trojan, worm or backdoor. 
Finally, in Section \ref{sec:allin}, we propose an original approach that performs two tasks: malware detection and ransomware detection. Our \textquote{bi-layered detection model} combines two optimized models for this and achieves slightly better results than benchmark models. This combined model provides an interesting flexibility capability to envision easier optimization and evolution steps in a production environment. We hope this paper could help improve cybersecurity solutions for malicious files analysis by providing new insights. 

\subsection{Future Works}
\label{sec:futur_works}

To accelerate and improve malware analysis activities, we plan to enrich results provided by ML and DL algorithms. In particular, it would be interesting to extend the method to other type of malware, and even to combine this method with the one we already proposed for the detection of packed and obfuscated files \cite{Marais2021}. We also plan to leverage Explainable Artificial Intelligence (XAI) and Interpretable Machine Learning (IML) concepts to provide decision support. In addition, we want to integrate clustering techniques and PE file behavioral analysis to improve malware classification results. Finally, we will have to assess the robustness of our different algorithms against adversarial attacks using for instance methods described in \cite{quertier2022merlin}.

\clearpage

\bibliographystyle{unsrt}
\bibliography{ref}

\end{document}